\documentclass[aps,prl,reprint,superscriptaddress]{revtex4-1}

\usepackage{graphicx}
\usepackage{dcolumn}
\usepackage{bm}

\begin{document}


\title{Observation of Terahertz Radiation via the Two-Color Laser Scheme with Uncommon Frequency Ratios}

\author{L.-L. Zhang}
\affiliation{Beijing Advanced Innovation Center for Imaging
Technology and Key Laboratory of Terahertz Optoelectronics (MoE),
Department of Physics, Capital Normal University, Beijing 100048,
China}
\author{W.-M. Wang}\email{hbwwm1@iphy.ac.cn or weiminwang1@126.com}
\affiliation{Beijing National Laboratory for Condensed Matter
Physics, Institute of Physics, CAS, Beijing 100190,
China}\affiliation{Beijing Advanced Innovation Center for Imaging
Technology and Key Laboratory of Terahertz Optoelectronics (MoE),
Department of Physics, Capital Normal University, Beijing 100048,
China}

\author{T. Wu}
\affiliation{Beijing Key Laboratory for
Precision Optoelectronic Measurement Instrument and Technology,
School of Optoelectronics, Beijing Institute of Technology, Beijing
100081, China}

\author{R. Zhang}
\affiliation{Beijing Key Laboratory for Precision Optoelectronic
Measurement Instrument and Technology, School of Optoelectronics,
Beijing Institute of Technology, Beijing 100081, China}

\author{S.-J. Zhang}
\affiliation{Beijing Key Laboratory for Precision Optoelectronic
Measurement Instrument and Technology, School of Optoelectronics,
Beijing Institute of Technology, Beijing 100081, China}

\author{C.-L. Zhang}
\affiliation{Beijing Advanced Innovation Center for Imaging
Technology and Key Laboratory of Terahertz Optoelectronics (MoE),
Department of Physics, Capital Normal University, Beijing 100048,
China}

\author{Y. Zhang}
\affiliation{Beijing Advanced Innovation Center for Imaging
Technology and Key Laboratory of Terahertz Optoelectronics (MoE),
Department of Physics, Capital Normal University, Beijing 100048,
China}

\author{Z.-M. Sheng}
\affiliation{SUPA, Department of Physics, University of Strathclyde,
Glasgow G4 0NG, United Kingdom} \affiliation{Key Laboratory for
Laser Plasmas (MoE) and School of Physics and Astronomy, Shanghai
Jiao Tong University, Shanghai 200240, China}\affiliation{IFSA
Collaborative Innovation Center, Shanghai Jiao Tong University,
Shanghai 200240, China}

\author{X.-C. Zhang}
\affiliation{The Institute of Optics, University of Rochester,
Rochester, New York 14627, USA} \affiliation{Beijing Advanced
Innovation Center for Imaging Technology and Key Laboratory of
Terahertz Optoelectronics (MoE), Department of Physics, Capital
Normal University, Beijing 100048, China}

\date{\today}

\begin{abstract}
In the widely-studied two-color laser scheme for terahertz (THz)
radiation from a gas, the frequency ratio of the two lasers is
usually fixed at $\omega_2/\omega_1=$1:2. We investigate THz
generation with uncommon frequency ratios. Our experiments show, for
the first time, efficient THz generation with new ratios of
$\omega_2/\omega_1=$1:4 and 2:3. We observe that the THz
polarization can be adjusted by rotating the longer-wavelength laser
polarization and the polarization adjustment becomes inefficient by
rotating the other laser polarization; the THz energy shows similar
scaling laws with different frequency ratios. These observations are
inconsistent with multi-wave mixing theory, but support the
gas-ionization model. This study pushes the development of the
two-color scheme and provides a new dimension to explore the
long-standing problem of the THz generation mechanism.
\end{abstract}


\maketitle

Terahertz (THz) waves have broad applications in THz spectroscopy
\cite{THz-spectroscopy,NonlinearTHz-spectroscopy} and THz-field
matter interactions \cite{THz-phy3,THz-CP}. These applications can
potentially benefit from powerful THz radiation sources with various
parameters via different laser-plasma-based schemes
\cite{Cook,Amico,Sheng,Gopal,Jin_prl}. For example, MV/cm-scale THz
radiation with either linear \cite{Cook,Xie,Kim} or elliptical
polarization \cite{WuHC,Dai_CP,Wen_CP,THz_PRL} can be generated from
gas plasma. THz radiation of near mJ can be produced via
relativistic laser interaction with solid plasma
\cite{Gopal,Liao1,Liao2,Jin}. Among these schemes, the two-color
laser scheme \cite{Cook} has been studied most widely
\cite{Kim2,THz_OE,YChen,Wang_TJ,Babushkin_prl,Babushkin_njp,
Kim_PRL,Sawtooth,Mechanism,YPChen_PRL} because it can provide
high-efficiency tabletop broadband sources. Generally, an 800nm pump
laser pulse passes through a frequency-doubling crystal to generate
a second-harmonic pulse and then the two pulses are mixed to produce
gas plasma. Up to now, the frequency ratio of the two-color pulses
has been always taken as $\omega_2/\omega_1=$1:2 in experiments,
although the fundamental-pulse wavelength longer than 800nm was
adopted in recent experiments to enhance the THz strength
\cite{WL_Scaling,Vvedenskii,Theory_WL} and the second-harmonic-pulse
frequency was detuned to yield ultra-broadband radiation
\cite{Thomson}. Since 2013 a few theoretical reports
\cite{THz_PRE,THz_w,THz_PRA} have predicted that the two-color
scheme could be extended to uncommon frequency ratios such as
$\omega_2/\omega_1=$1:4, 2:3, but these predictions have not yet
been verified experimentally.

In this Letter, we present the first experimental demonstration of
THz generation with uncommon frequency ratios. With the
$\omega_1$-laser wavelength fixed at 800nm and 400nm, respectively,
a scan of the $\omega_2$-laser wavelength from 1200nm to 1600nm
shows that the THz energies have three resonantlike peaks located
near $\omega_2/\omega_1=$ 1:4, 1:2, and 2:3. The energies at these
peaks are at the same order. Beyond the previous predictions
\cite{THz_PRE,THz_w,THz_PRA}, we find that the THz polarization can
be adjusted by rotating the $\omega_2$-pulse polarization and
however, the polarization adjustment becomes inefficient by rotating
the $\omega_1$-pulse polarization. In this Letter we define the
$\omega_1$ pulse as the higher-frequency one. These observations
agree with our particle-in-cell (PIC) simulations and a model based
on field ionization.

The current experiments with the new frequency ratios also provide a
new dimension to explore further the THz-generation mechanism. Since
2000 it has been a frequently-discussed topic: whether this THz
generation can be attributed to multi-wave mixing
\cite{Cook,Xie,xcZhang2016}, field ionization
\cite{Kim,THz_OE,ZhangDW}, or to both \cite{Theory_WL,Mechanism}.
First, multi-wave mixing theory predicts that the THz energy
$\varepsilon_{THz}$ scales with $(P_1) (P_2)^2$ in the original
scheme, where $P_1$ and $P_2$ are powers of the two pulses. With
$\omega_2/\omega_1=1:4$ and $2:3$, $\varepsilon_{THz}$ should follow
different scaling laws $(P_1) (P_2)^4$ and $(P_1)^2 (P_2)^3$,
respectively. In the experiments we observe complex dependence of
$\varepsilon_{THz}$ on $P_1$ and $P_2$ similar with different
$\omega_2/\omega_1$, in disagreement with these scaling laws.
Second, we observe that the THz polarization varies only with
rotating the polarization of the longer-wavelength laser, which is
inconsistent with the symmetric nature in the susceptibility tensor
required by the multi-wave mixing theory \cite{Xie}.

\begin{figure}[htbp]
\includegraphics[width=3.2in]{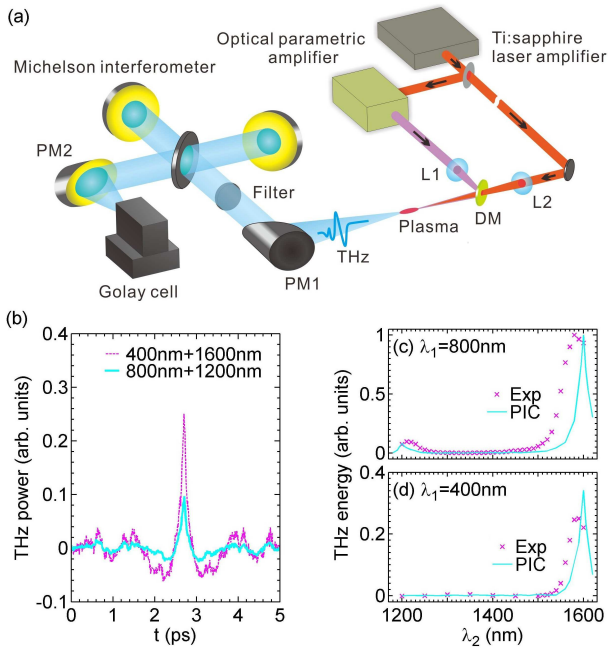}
\caption{\label{fig:epsart}(a) Experimental setup: L, lens; DM,
dichromatic mirror; PM, parabolic mirrors. (b) THz waveforms with
$\omega_2/\omega_1=1:4$ and $2:3$, respectively, obtained from the
autocorrelation measurements, in which the THz powers are normalized
by the one with the 800nm and 1600nm pulses. (c), (d) THz energy as
a function of the second pulse wavelength $\lambda_2$, where the
first pulse wavelength $\lambda_1$ is fixed as 800nm in (c) and
400nm in (d). Powers of the two pulses are taken as $P_1=$ 120mW and
$P_2=$ 400mW in (c) and $P_1=$ 180mW and $P_2=$ 250mW in (d).}
\end{figure}

\emph{Experimental setup.}$-$ Figure 1(a) shows a schematic of our
experiment. The laser pulse from a Ti:Sapphire amplifier (Spitfire,
Spectra Physics) with a central wavelength of 800 nm, duration of 50
fs, and repetition rate of 1 kHz. The pulse with total energy of 5.3
mJ is split into two parts. The part with 3.5 mJ is used to pump an
optical parametric amplifier (TOPAS), which delivers a pulse
wavelength tunable from 1200 nm to 1600 nm (the $\omega_2$ pulse).
The remaining energy is used as the $\omega_1$ pulse of 800 nm
wavelength [see Fig. 1(c) as an example]. In another group of
experiments [see Fig. 1(d)], the 800 nm pulse passes through a
switchable $\beta$-barium borate (BBO) crystal and band-pass filter
to generate 400nm-wavelength pulse (the $\omega_1$ one). The
$\omega_1$ and $\omega_2$ pulses propagate collinearly using a
dichromatic mirror and have a confocal spot focused by two convex
lenses with equal focal length f=12.5cm. Both pulses are linearly
polarized in the horizontal plane initially and their polarizations
can be independently controlled by half-wave plates. Powers can also
be independently adjusted through optical attenuators. The two
pulses irradiate air and produce a few millimeter of plasma.

We use an off-axis parabolic mirror to collect and collimate the
forward THz radiation generated from the gas plasma after
eliminating the pump laser pulses with a long-pass THz filter (Tydex
Ltd.). To measure the horizontal and vertical components of the
radiation, a wire grid polarizer is employed. A Golay THz detector
with a 6 mm diameter diamond input window (Microtech SN:220712-D) is
used to measure the radiation energy, where the detector shows a
nearly flat response in the spectral range from 0.1 THz to 150 THz.
The voltage signal is fed into a lock-in amplifier referenced to a
15 Hz modulation frequency. To obtain the THz radiation bandwidth,
autocorrelation measurement is carried out by a Michelson
interferometer containing a silicon wafer.

\begin{figure}[htbp]
\includegraphics[width=3.2in]{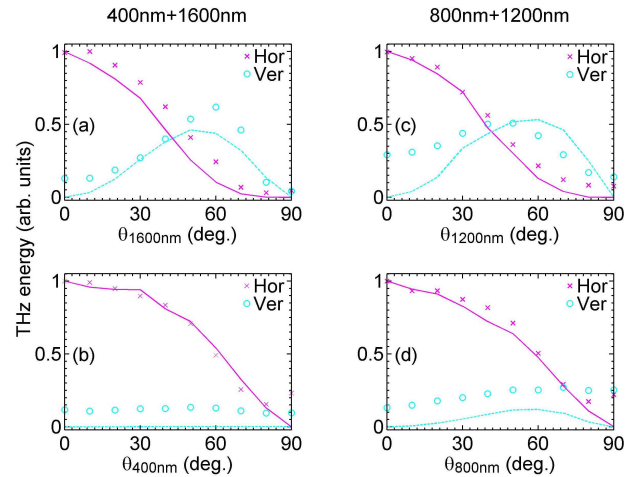}
\caption{\label{fig:epsart}THz energies of the horizontal and
vertical components as a function of the rotation angle $\theta$ of
the field polarization of (a) the 1600nm pulse, (b) 400nm pulse, (c)
1200nm pulse, and (d) 800nm pulse, respectively, where when
polarization of one pulse is rotated, polarization of the other
pulse is fixed at the horizontal. Experimental results are shown by
crosses and circles and PIC results by lines. The left column
corresponds to the case with the 400 nm (with 180mW) and 1600nm
(250mW) pulses and the right to the case with the 800nm (120mW) and
1200nm (400mW) pulses.}
\end{figure}

\begin{figure}[htbp]
\includegraphics[width=3.2in]{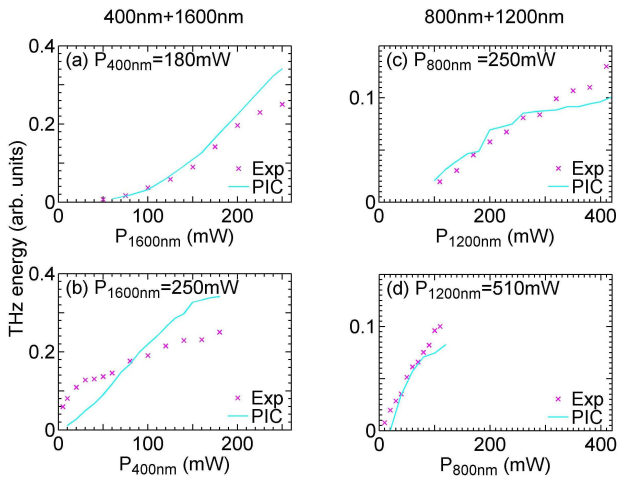}
\caption{\label{fig:epsart}THz energy as a function of the power of
(a) the 1600nm pulse, (b) 400nm pulse, (c) 1200nm pulse, and (d)
800nm pulse, respectively, where when the power of one pulse is
changed, the power of the other pulse is fixed. The left column
corresponds to the case with the 400nm and 1600nm pulses and the
right to the case with the 800nm and 1200nm pulses.}
\end{figure}

\emph{Experimental and PIC-simulation results.}$-$ We first present
the experimental and PIC simulation results and then explain them
with a theoretical model based on field ionization of gas. In our
experiments, we observe dependence of THz generation upon
$\omega_2/\omega_1$, laser polarization, and laser power,
respectively, as shown in Figs. 1(b)-3. In these figures except Fig.
1(b), our PIC simulation results are also shown. First, the measured
THz waveforms plotted in Fig. 1(b) show that the THz peak powers
with $\omega_2/\omega_1=1:4$ and $2:3$ are about 30\% and 10\%
compared with $\omega_2/\omega_1=1:2$. By scanning $\omega_2$ from
1200 nm to 1600 nm, we observe that the THz radiation can be
effectively generated only around $\omega_2/\omega_1=2:3$ and $1:2$
in Fig. 1(c) with the $\omega_1$ pulse of 800 nm as well as around
$\omega_2/\omega_1=1:4$ in Fig. 1(d) with the $\omega_1$ pulse of
400 nm. Note that these observed THz peaks have small shifts
($10-20$ nm in wavelength) from the ones exactly at
$\omega_2/\omega_1=1:4$, $2:3$, and $1:2$ obtained in the PIC
results, which could be caused by inaccuracy of laser wavelengths
output from TOPAS. Second, we observe in Fig. 2 that the THz
polarization can be adjusted by rotating the polarization of the
$\omega_2$ (longer-wavelength) pulse, but the polarization
adjustment becomes inefficient by rotating the $\omega_1$-pulse
polarization. This phenomenon is observed in all the cases of
$\omega_2/\omega_1=1:4$ [Figs. 2(a) and 2(b)],
$\omega_2/\omega_1=2:3$ [Figs. 2(c) and 2(d)], and
$\omega_2/\omega_1=1:2$. For example in the case
$\omega_2/\omega_1=1:4$, when the 1600nm-pulse polarization is
rotated from the horizontal to the vertical in Fig. 2(a), the THz
horizontal component is weakened continuously and the vertical
component is first strengthened and then weakened, as observed in
previous experiments \cite{Vvedenskii} with $\omega_2/\omega_1=1:2$.
However, when the 400nm-pulse polarization is rotated in Fig. 2(b),
the THz vertical component is kept at a low level similar to that at
$\theta=0$ and $90^o$, which is expected to be at noise level. These
observations are reproduced by our PIC simulations. Third, the
dependence of the THz energy upon the laser powers does not obey the
scaling laws predicted by the multi-wave mixing theory, as seen in
Fig. 3. The curves in this figure illustrate complex dependence in
both cases $\omega_2/\omega_1=1:4$ and $2:3$ and each curve in the
starting phase appears as a linear dependence, in reasonable
agreement with the PIC results.

The agreement between the PIC (near-field radiation) and
experimental results (far-field radiation) suggests that the
far-field radiation observed should be mainly contributed from a
short gas-plasma zone in which the pulses have the highest
intensities, as modeled in our PIC simulations. In our PIC
simulations, we employ a 0.6-millimeter-long nitrogen gas to save
computational time. We adopt the same laser parameters as in the
experiments and assume that on the front-end of this gas the laser
pulses just reach the highest intensities (at the order of
$10^{14}~\rm W/cm^2$) and have the spot radius of 50 $\mu m$. Our
PIC simulations are performed with the KLAPS code \cite{KLAPS}, in
which field ionization of gas is realized by Monte Carlo method,
movement of the created electrons is computed by the relativistic
motion equation, and a full-Maxwell-equation solver is included to
calculate generation and propagation of both lasers and radiation.
It can self-consistently compute plasma production and net current
formation via laser-field ionization, dynamics of the net current in
the plasma, and THz generation. Therefore, our PIC simulation can
give near-field THz radiation with very few approximations. Note
that the far-field radiation is expected to be composed of all
near-field sources \cite{Babushkin_prl,Babushkin_njp,YPChen_PRL} and
a simplified near-field model was used to well explain THz
generation experiments in Ref. \cite{Vvedenskii}.



\emph{Theoretical model.}$-$ To interpret the PIC results and the
experiment results, we present theoretical analysis based on a net
or transient current model. It was first proposed by Kim \emph{et
al.} \cite{Kim,Kim2} to show current formation due to asymmetric
field ionization. Then, Wang \emph{et al.} proposed a near-field
model including the current dynamics in plasma
\cite{THz_OE,THz_PoP,THz_PRL}. The THz radiation generation includes
two processes: net-current formation via field ionization and THz
generation as the current is modulated by the plasma. The former
lasts a time shorter than the laser duration 50 fs while the latter
has a timescale at the THz period about 1 ps. Therefore, one can
calculate the two processes respectively. The net current
$\mathbf{J}_{0}=-en_e\mathbf{v}_{0}$ can be given by
\begin{eqnarray}
\mathbf{J}_{0}=\frac{e^2n_e \mathbf{A}_L(\psi_0)}{m_ec},
\end{eqnarray}
where $\mathbf{v}_{0}=-e \mathbf{A}_L(\psi_0)/m_ec$, $\mathbf{A}_L$
is the laser vector potential, $\psi=t-z/c$, and $\psi_0$ is the
position where electrons are created. Note that nearly all electrons
are periodically created at the same relative position in different
periods of the laser fields in the cases $\omega_2/\omega_1=$ 1:4,
1:2, 2:3, respectively, as shown in Ref. \cite{THz_PRA}. The
electron density is given according to $\partial n_e/\partial
t=(n_a-n_e)w(E_L)$, where $w(E_L)$ is the ionization rate
\cite{ADK,ADK_revised,MultiPho_ADK} in the laser field amplitude
$E_L$ and $n_e$ and $n_a$ are the electron and initial atom
densities, respectively. After passage of the laser pulses, the
generated radiation interacts with the current, the electron
velocity becomes $\mathbf{v}=\mathbf{v}_{0}+e\mathbf{A}_{THz}/m_ec$,
and consequently the current turns to
$\mathbf{J}=\mathbf{J}_{0}-e^2n_e\mathbf{A}_{THz}/m_ec$, where
impacts of the radiation ponderomotive force on $n_e$ can be
ignored. Then, the THz radiation can be described by
\begin{eqnarray}
\left[\nabla^2-\frac{1}{c^2}\frac{\partial^2}{\partial
t^2}-\frac{\omega_p^2}{c^2}\right]\mathbf{A}_{THz}=-4\pi\mathbf{J}_{0}/c,
\end{eqnarray}
where $\omega_p=\sqrt{4\pi e^2 n_e/m_e}$ is the plasma oscillation
frequency. Equation (2) is difficult to analytically solve since the
pulse length of the THz radiation is longer than the spot size
($\sim50 \rm \mu m$) and a one-dimensional approximation
\cite{THz_PoP} cannot be taken. In the following, we will show that
numerical calculation of Eq. (1) and simple analysis of Eq. (2) can
explain the experimental results presented previously.

\emph{Dependence on laser frequency ratio.}$-$ From Eqs. (1) and
(2), one can obtain the THz amplitude $A_{THz} \propto J_{0} \propto
A_L(\psi_0)$. Peaks of THz energies appear at peaks of
$A_L(\psi_0)$. Our calculation shows three resonance-like peaks of
$A_L(\psi_0)$ located at $\omega_2/\omega_1=$ 1:4, 1:2, 2:3. To
quantitatively compare the THz energies at the three peaks, we also
calculate $J_{0}$ which depends on both $A_L(\psi_0)$ and $n_e$.
Calculating $J_{0}$ by Eq. (1) gives the values of $J_{0}$ as
$0.29:1:(-0.58)$. Then, the THz energies are $0.084:1:0.34$, which
is in agreement with the experimental results of $0.097:1:0.26$ as
seen in Figs. 1(c) and 1(d).

\emph{Dependence on laser polarization.}$-$ According to Eqs. (1)
and (2), the THz radiation should have only the $x$ component if the
two pulses have the same polarization along the $x$ direction. Once
the polarization of one pulse is rotated to have the $y$ component,
the radiation could have both $x$ and $y$ components. We take the
laser electric fields as $E_{L,x}=f(\psi)[a_1 \sin(\omega_1 \psi) +
a_2\cos(\theta) \sin(\omega_2 \psi)]$ and $E_{L,y}=f(\psi)
a_2\sin(\theta) \sin(\omega_2 \psi)$, where $\theta$ is the rotation
angle and $f(\psi)$ is the envelope profile. The vector potential
can be written by $A_{L,x}=cf(\psi)[a_1 \cos(\omega_1 \psi)/\omega_1
+ a_2\cos(\theta) \cos(\omega_2 \psi)/\omega_2]$ and
$A_{L,y}=cf(\psi) a_2\sin(\theta) \cos(\omega_2 \psi)/\omega_2$
since $\partial f(\psi)/\partial \psi \ll \omega_1$ and $\omega_2$
for the pulse duration of 50 fs. Electrons are created at the
maximum of $E_L^2=f^2(\psi)[a_1^2 \sin^2(\omega_1 \psi) + a_2^2
\sin^2(\omega_2 \psi) +2 a_1 a_2 \cos(\theta) \sin(\omega_1 \psi)
\sin(\omega_2 \psi)]$, i.e., at $\frac{\partial |E_L|}{\partial
\psi}=0$, which gives $\omega_2 \psi_0=1.937$ for $\theta=0$ ($a_1$
and $a_2$ are computed from $P_{400nm}=180$ mW and $P_{1600nm}=250$
mW, respectively). Our calculation shows that $\psi_0$ varies
slightly with the change in $\theta$, because $\sin(\omega_1 \psi)$
and $\sin(\omega_2 \psi)$ are close to 1 to produce the maximum of
$|E_L|$ with $a_1\sim a_2$. With $\omega_2 \psi_0=1.937$, $\partial
[\frac{\partial |E_L|}{\partial \psi}]/\partial
[\cos(\theta)]\simeq0.06$ can be derived, which suggests that when
$\cos(\theta)$ is changed from 1 to 0 ($\theta$ from 0 to $\pi/2$),
$\frac{\partial |E_L|}{\partial \psi}|_{\psi_0+\epsilon}=0$ is
always satisfied if $\psi_0$ is shifted by a small value $\epsilon$.

Therefore, both $|E_L(\psi_0)|$ and $|A_{L,x}(\psi_0)|$ decreases as
$\theta$ is increased from 0 to $\pi/2$, where $A_{L,x}(\psi_0)<0$
and $\cos(\omega_2\psi_0)<0$. Decrease of $|E_L(\psi_0)|$ and
$|A_{L,x}(\psi_0)|$ leads to a reduction of ionization rates and net
velocities of electrons, respectively, which can explain the
weakening THz horizontal (or $x$) component with $\theta$ in Fig.
2(a). This figure also shows that the vertical component is first
strengthened from zero and then weakened, which is caused by the
increasing $|A_{L,y}(\psi_0)|$ and decreasing $|E_L(\psi_0)|$ with
$\theta$. The peak of the vertical component is observed about
$\theta=60^o$ approaching the PIC result. Our simulations show the
optimized $\theta$ within $40^o-70^o$ dependent of the laser
intensities and frequencies, determined by the balancing point of
the increasing $|A_{L,y}(\psi_0)|$ and the decreasing
$|E_L(\psi_0)|$.

In Fig. 2(b) the 400 nm pulse polarization (the $\omega_1$ pulse) is
rotated, the THz vertical component is kept at a low level (noise
level in the experiments and near zero in the PIC simulations).
Rotating the $\omega_1$ or $\omega_2$ pulse, $|E_L|$ is unchanged
and consequently, $\frac{\partial |E_L|}{\partial \psi}=0$ gives the
same $\omega_2 \psi_0=1.937$ for $\theta=0$ and $\psi_0$ varies
slightly with $\theta$. Therefore, the horizontal component in Fig.
2(b) shows the similar dependence to Fig. 2(a) for the same reason
addressed previously. However, the vertical component depends
strongly on the laser frequency. When rotating the $\omega_1$ pulse,
$A_{L,y}^{\omega_1}(\psi_0)=cf(\psi_0) a_1\sin(\theta) \cos(\omega_1
\psi_0)/\omega_1$. While rotating the $\omega_2$ pulse,
$A_{L,y}^{\omega_2}(\psi_0)=cf(\psi_0) a_2\sin(\theta) \cos(\omega_2
\psi_0)/\omega_2$. One can obtain
\begin{eqnarray}
\frac{A_{L,y}^{\omega_1}(\psi_0)}{A_{L,y}^{\omega_2}(\psi_0)}\simeq
-(\frac{\omega_2}{\omega_1})^2=-(\frac{\lambda_1}{\lambda_2})^2,
\end{eqnarray}
where we have used
$a_1\omega_1\cos(\omega_1\psi_0)=-a_2\omega_2\cos(\omega_2\psi_0)$
derived from $\frac{\partial |E_L|}{\partial \psi}=0$ with
$\theta=0$ since $\psi_0$ slightly depends upon $\theta$. According
to Eq. (3), the THz energy of the vertical component is decreased to
$1/256\simeq0.004$ when the rotated pulse is changed from the
$\omega_2$ one to the $\omega_1$ with $\omega_2/\omega_1=1:4$; and
the THz energy is decreased to $16/81\simeq0.2$ with
$\omega_2/\omega_1=2:3$. These are in good agreement with our PIC
results as shown in Figs. 2(b) and 2(d). Since such low levels of
THz energies cannot be resolved in our experiments, the vertical
component is observed to be nearly unchanged with varying $\theta$.
Similar results are also observed in our experiments when the 800 nm
and 1600 nm pulses are used.

Note that the observed THz polarization dependence is inconsistent
with the multi-wave mixing model \cite{Xie}. For example with
$\omega_2/\omega_1=1:4$, the fifth-order susceptibility tensor
$\chi$ for THz generation has $\chi^{x}_{xyyyy}=\chi^{y}_{yxxxx}$
because of the symmetry, where the  superscript of $\chi$ represents
the THz polarization and the subscripts represent the polarization
of the $\omega_1$ wave and the four $\omega_2$ waves, respectively.
$\chi^{x}_{xyyyy}=\chi^{y}_{yxxxx}$ requires that the horizontal THz
component in Fig. 2(a) should have the same level as the vertical
THz component in Fig. 2(b). In contrast, Figs. 2(a) and 2(b) gives
$\chi^{x}_{xyyyy}\gg\chi^{y}_{yxxxx}$. Besides, both our PIC and
experimental results show obvious differences from $\cos^2(\theta)$
scaling for the horizontal component and $\sin(2\theta)$ for the
vertical component, which was derived under the different condition
$a_1\ll a_2$ and with $\omega_2/\omega_1=1:2$ \cite{Vvedenskii}.

\emph{Dependence on laser power.}$-$ Figure 3 shows complex
dependence of the THz energy on the laser power for
$\omega_2/\omega_1=1:4$ and $2:3$, which significantly deviates from
the scaling of $(P_1) (P_2)^4$ and $(P_1)^3 (P_2)^2$ predicted by
the multi-wave mixing theory. This can be attributed to complex
dependence of the ionization rates on the laser intensities since
the intensities span one to two orders of magnitude, which adds
significant complexity to theoretical analysis. The analysis becomes
simpler when the power of one pulse is changed in a low level within
[$P_a,P_b$] and the power of the other pulse is fixed at a much
higher value $P_c$ ($P_c\gg P_b$), where the ionization rate and the
ionization position $\psi_0$ vary slightly. This is the case in the
starting stage in each curve in Fig. 3. According to $\frac{\partial
|E_L|}{\partial \psi}(\psi_0)=0$ for the two pulses with the same
polarization, one can obtain
$A_{L,x}(\psi_0)=a_1cf(\psi_0)\cos(\omega_1 \psi_0)[1/\omega_1-
\omega_1/\omega_2^2]$ or $A_{L,x}(\psi_0)=a_2cf(\psi_0)\cos(\omega_2
\psi_0)[1/\omega_2- \omega_2/\omega_1^2]$. In the case with the
laser powers $P_1\gg P_2$ and $P_1\ll P_2$, $\psi_0$ varies slightly
with the change of $a_1$ and $a_2$ and therefore,
$|A_{L,x}(\psi_0)|$ is linearly proportional to $a_1$ or $a_2$,
i.e., the THz energy is linearly proportional to $P_1$ or $P_2$.
This linear dependence is observed within the starting stage in each
curve in Fig. 3 with either $\omega_2/\omega_1=1:4$ or $2:3$ (one
can also observe similar results in previous experiments with
$\omega_2/\omega_1=1:2$ \cite{Vvedenskii}). Note that the PIC and
experimental results are not in precise agreement. In the PIC
simulations we assume that the laser pulses with different powers
have the same spot radius of 50 $\mu m$ when they reach the highest
intensities. However, the spot radius will depend on the power,
unfortunately, exploration of this complex dependence is beyond the
scope of this work.

In summary, we have experimentally shown that the two-color scheme
can still work when $\omega_2/\omega_1$ of 1:2 is changed to 1:4 and
2:3. The THz polarization can be adjusted more efficiently by
rotating the polarization of the longer-wavelength pulse from the
horizontal to the vertical because the THz vertical component
follows a fourth-power law of the laser wavelength, which is
inconsistent with the multi-wave mixing theory. We have observed a
complex dependence of the THz energy when the power of one of the
two pulses is varied over a large range. A linear dependence with
different $\omega_2/\omega_1$ has also been observed when the power
of one pulse is varied within a limited range much lower than the
power of the other pulse. These dependencies disagree with the
scaling laws given by the multi-wave mixing theory. These
observations have been well explained by our PIC simulations and a
model based on field ionization.

\begin{acknowledgments}
This work was supported by the National Basic Research Program of
China (Grant No. 2014CB339806 and No. 2014CB339801), the National
Natural Science Foundation of China (Grants No. 11375261, No.
11775302, and No. 11374007), Science Challenge Project of China
(Grant No. TZ2016005), and the Strategic Priority Research Program
of the Chinese Academy of Sciences (Grants No. XDB16010200 and
XDB07030300).
\end{acknowledgments}

\end{document}